# A Soft Processor Overlay with Tightly-coupled FPGA Accelerator

Ho-Cheung Ng, Cheng Liu, Hayden Kwok-Hay So
Department of Electrical & Electronic Engineering, The University of Hong Kong
{hcng, liucheng, hso}@eee.hku.hk

*Abstract*—FPGA overlays are commonly implemented as coarse-grained reconfigurable architectures with a goal to improve designers' productivity through balancing flexibility and ease of configuration of the underlying fabric. To truly facilitate full application acceleration, it is often necessary to also include a highly efficient processor that integrates and collaborates with the accelerators while maintaining the benefits of being implemented within the same overlay framework.

This paper presents an open-source soft processor that is designed to tightly-couple with FPGA accelerators as part of an overlay framework. RISC-V is chosen as the instruction set for its openness and portability, and the soft processor is designed as a 4-stage pipeline to balance resource consumption and performance when implemented on FPGAs. The processor is generically implemented so as to promote design portability and compatibility across different FPGA platforms.

Experimental results show that integrated software-hardware applications using the proposed tightly-coupled architecture achieve comparable performance as hardware-only accelerators while the proposed architecture provides additional run-time flexibility. The processor has been synthesized to both low-end and high-performance FPGA families from different vendors, achieving the highest frequency of 268.67 MHz and resource consumption comparable to existing RISC-V designs.

## I. INTRODUCTION

By raising the abstraction level of the underlying configurable fabric, many early works have already demonstrated the promise of using FPGA overlays to improve designer's productivity in developing hardware accelerators [1], [2]. While such hardware accelerators can often deliver significant performance improvement over their software counterparts, they are often fixed in functionality and lack the flexibility to process irregular input or data that depends on run-time dynamics. To truly take advantage of the performance benefit of hardware accelerators, it is therefore desirable to have an efficient CPU in the overlay tightly-coupled with the accelerator to control its operations and to maintain compatibility with the rest of the software system.

To illustrate these intricate hardware-software codesign challenges, Algorithm 1 shows a simple design that accelerates the Sobel edge detection algorithm in such heterogeneous system. In this implementation, an accelerator that computes $16 \times 16$ output pixels at a time is implemented in FPGA. During run time, depending on the user input image size, the software reuses this hardware accelerator for as many *complete* $16 \times 16$ output pixels as possible. The remaining odd pixels, as well as pixels on the boundary of the image where the standard filter kernel cannot readily operate on, are handled in software.

```
Data: Pixels of size N × N
1  # define BUF 16 // HW computes 16x16 output pixels
2  for r := 0 to N − 1 do
3      for c := 0 to N − 1 do
4          if pixel[r, c] is edge then
5              SW_SOBEL( pixel, r, c );
6          else if ((r − 1) % BUF) == 0 &&
7                  (c − 1) % BUF) == 0 then
8              HW_SOBEL( pixel, r, c );
9          else
10             continue;
11     end
12 end
13 end
```

**Algorithm 1:** Pseudocode for Sobel edge detector. As the hardware accelerator operates on a fixed $16 \times 16$ array of output pixel at a time, software passes control to the accelerator only for cases when all $17 \times 17$ pixels are available. Otherwise, the computation is carried out in software. Assume $N - 2$ is a multiple of `BUF`.

While the design of Algorithm 1 may be specific to the particular implementation of Sobel edge detection, it highlights several challenges commonly faced by many real-world hardware-software designers. First of all, because of the limited flexibility of most hardware accelerators, the controlling software must ensure the necessary input data are available before the accelerator is launched. Furthermore, unless the hardware accelerator is arbitrarily flexible, software running in the CPU must also be able to process any run time data that cannot readily be processed by the accelerator.

In view of the above, this paper proposes the use of a *small, open source* soft processor to provide fine-grained control for the hardware accelerator in the context of an overlay framework. The core is designed to be *tightly-coupled* with the hardware accelerator in order to minimize the overhead involved with switching control between hardware and software. RISC-V `RV32I` [3] is chosen as the ISA for its openness and simplicity. Finally, the core is generically designed in order to promote design portability and compatibility.

As such, we consider the main contribution of this work rests on the demonstration of the benefits of tightly-coupling a lightweight CPU with hardware accelerator to serve within a combined overlay architecture. We also demonstrated that a simple RISC-V CPU with 4-stage pipeline is adequate to provide control while delivering reasonable performance and maintaining software compatibility.

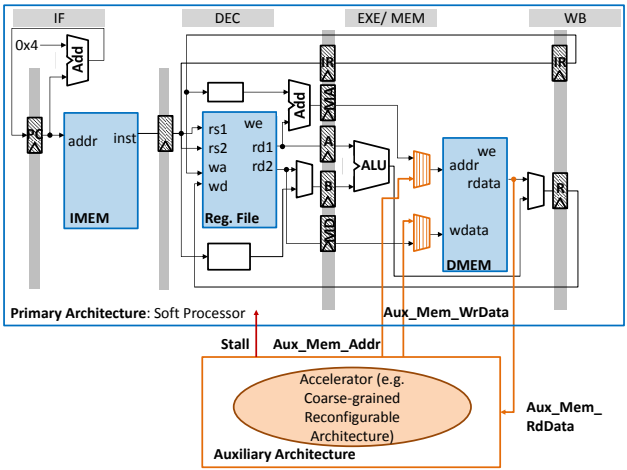

Fig. 1: High-level diagram of the proposed tightly-coupled architecture.

TABLE I: The format of BAA instruction.

| 31   20 | 19  15 | 14  12 | 11  7 | 6    0 |
|---|---|---|---|---|
| imm[11:0] | rs1 | funct3 | – | opcode |
| 12 | 5 | 3 | 5 | 7 |
| offset[11:0] | base | width | – | BAA |

TABLE II: The format of RPA instruction.

| 31   20 | 19  15 | 14  12 | 11  7 | 6    0 |
|---|---|---|---|---|
| imm[11:0] | rs1 | funct3 | – | opcode |
| 12 | 5 | 3 | 5 | 7 |
| offset[11:0] | base | width | – | RPA |

In the next section, we elaborate on the design and implementation of the soft processor and the tightly-coupled architecture. We then evaluate the performance of the proposed architecture in Section III and discuss related work in Section IV. We make conclusions in Section V.

## II. Design and Implementation Details

Figure 1 displays a high-level diagram of the proposed tightly-coupled architecture. This design consists of two entities where an accelerator can be integrated with the single-issue, in-order processor pipeline by sharing the data memory (DMEM). To ensure correct execution flow, control signal is fed from the accelerator to the processor control path so that the pipeline is stalled correctly when the accelerator is carrying out execution.

### A. The Soft Processor

In order to reduce substantial resource consumption while maintaining certain efficiency, the soft processor is designed as a 4-stage pipeline by integrating the execution-stage and memory-stage together. This eliminates the load-use hazard where a LOAD instruction is followed by an instruction that uses the value which is just transferred from DMEM to Register File.

In spite of the above advantage, since the memory-stage is now merged with the execution-stage and the memory address needs to be ready before the load/store instruction reaches DMEM, an extra 32-bit adder is placed at the end of the decode-stage. This could incur extra resource consumption and additional pipeline delay.

We found that, from the synthesis results, the proposed 4-stage processor consumes 18% fewer amounts of FPGA registers and LUTs when compared with the traditional 5-stage pipeline. We believe that this reduction is important for portability and compatibility concerns, especially when the soft processor is ported onto the legacy FPGA devices which could be intrinsically small in size. Also, the additional delay incurred by the extra adder could be partly compensated by the two additional multiplexers which are placed in front of DMEM.

Moreover, as the benefit of using a virtual layer of overlay architecture on FPGA rests on improving designer's productivity while providing certain customization capabilities, the proposed soft processor can also be customized in terms of the IMEM and DMEM size. Developers can change the memory sizes by modifying a few lines of macro or execute a program that comes along with the soft processor design package.

It is important to note that, in order to further minimize FPGA resource consumptions, components that are not strictly necessary for processor overlay execution are removed from the current implementation. These components include the Control Status Registers (CSR) and their corresponding logic. Future versions of the processor implementation will incorporate the CSR back and will provide tools to allow developers to remove them during the processor customization.

### B. The Tightly-coupled Architecture

In order to support direct memory access and allow zero-overhead transfers of control, the soft processor is tightly-coupled with the hardware accelerator as illustrated in Figure 1. In the proposed framework, Multiple Runtime Architecture Computer(MURAC) model [4] is adopted to handle the transfer of control when the execution is switched from one architecture to another.

The execution model of the tightly-coupled architecture follows naturally from the MURAC model where the proposed 4-stage pipeline is defined as the *primary architecture* while the hardware accelerator is defined as the *auxiliary architectures*. Switching between these two architectures in runtime is achieved by the Branch-Auxiliary Architecture (BAA) and Return-To-Primary-Architecture (RPA) machine instructions. Logically, a MURAC machine contains only one address space that is shared between two architectures.

*1) Custom Instruction-set Extension:* To apply MURAC machine for the proposed tightly-coupled architecture, custom instruction-set is used to implement the BAA and RPA instructions.

Custom instruction-set extension [5] is an important feature in RISC-V RV32I in the sense that it provides opportunity for designers to integrate other hardware modules such as accelerators onto a standard RISC-V processor. It also provides a unified programming model along various and future RISC-V processor designs, which makes it easier to leverage software development efforts for the ISA toolchain during the processor customization process.

*2) BAA and PRA Instructions Format:* In the proposed architecture, opcode space *custom-0* is selected to implement the `BAA` and `RPA` instructions. The format of this opcode is defined to be `inst[6:0]==0001011`. For the benefits of regularity and simplicity of the decoding hardware, both instructions follow the format of I-type instruction.

Table I displays the the format of `BAA` instruction which resembles the format as in `LOAD` instruction. The fields `base` and `offset` are added together to form a memory address location. It represents an address location that points to an array of data. This address is passed to the auxiliary architecture, i.e. accelerator during the execution of the `BBA` instruction.

The array passed to the auxiliary architecture acts like the command line arguments in any `C/C++` program. It stores up the data that is needed by accelerator. The first data represents the number of elements inside the array.

The field `width`, on the other hand, is used to distinguish between the `BAA` and `RPA` instructions, as they share similar encoding. The format of the `RPA` instruction is shown in Table II.

The fields `base` and `offset` in `RPA` instruction are added together to form a return memory address. This instruction acts like a return instruction where the program is unconditionally jumped to `base+offset`. Currently, the processor will branch to the address (`PC+4`) after the auxiliary architecture finishes its execution.

*3) Modifications of the Soft Processor:* In the control-path, extra control registers are defined to decode the `BAA` and `RPA` instructions. These registers are used to recognize the custom instruction with the help of comparators. In addition, stall signals are fed from the auxiliary architecture to the control-path so that the processor would not be executing once the control is passed to accelerator.

On the other hand, when the execution is branched to the auxiliary architecture, the processor pipeline is stalled and components on the data-path such as `DMEM` would not be accessed. This makes resource sharing between two architectures possible. In the proposed architecture, `DMEM` is designed to be shared with the accelerator. A number of multiplexers are added before the inputs of the `DMEM` and the output of `DMEM` is also fed to the auxiliary architecture. The control path would assert the correct `sel` signal for the multiplexers when the control is branched to auxiliary architecture.

## III. EXPERIMENTS & MEASUREMENTS

To study the design implications of the proposed framework, a set of real-life applications was programmed and run on the tightly-coupled architecture. In addition, the soft processor was also benchmarked and compared with an existing similar

TABLE III: Parameters and configurations of the benchmarks.

| MM | FIR | KM | SE |
|---|---|---|---|
| Matrix Size | # of Input/ # of Taps+1 | # of Nodes/ Centroids/ Dimension | # of Vertical Pixels/ # of Horizontal Pixels |
| 100×100 | 10000/50 | 5000/4/2 | (128+2)/(128+2) |

TABLE IV: The loop kernels were unrolled by the following factors before transferring to the auxiliary architecture for acceleration.

|  | MM | FIR | KM | SE |
|---|---|---|---|---|
| Full Loop | 100×100×100 | 10000×50 | 5000×4×2 | (128+2)×(128+2)×3×3 |
| Unrolling | 1×5×100 | 50×50 | 125×4×2 | 16×16×3×3 |

RSIC-V design. Finally, resource consumption and design portability of the soft processor were evaluated to warrant the merit of the proposed framework as FPGA overlay.

### A. Evaluation of the Tightly-coupled Architecture

*1) Experimental Setup:* Four real-life applications including matrix-matrix multiplication (MM), finite impulse response (FIR) filter, K-mean clustering algorithm (KM) and Sobel edge detector (SE) were taken as the benchmark to evaluate the tightly-coupled architecture. These applications are essentially loop kernels which are highly parallelizable and can be mapped to FPGA for performance acceleration. The parameters of the benchmark are listed in Table III.

To understand the underlying influence of the tightly-coupled architecture (TIGHTLY-COUPLED), the inner loop nest in each application was partially unrolled. Detailed loop unrolling configurations can be found in Table IV. The unrolled loop body was executed on a handcrafted hardware design (auxiliary architecture) for acceleration while miscellaneous loop controls and the boundary conditions that could not be covered by the accelerators were executed on the soft processor (primary architecture).

Meanwhile, we also provided a pure hardware implementation (HW) and a pure software implementation (SW) for each of the applications for comparison. Basically, HW had the whole application implemented on FPGA with handcrafted hardware design. It can process not only the loop body but also the loop control as well as the boundary condition. Therefore, the execution can be done entirely on FPGA without the interventions from the soft processor. SW simply ignored the accelerator and had the whole application running on the proposed soft processor.

Finally, we assumed that every piece of data was already cached in DMEM so as to eliminate the influence from the memory and maximize the impact of the soft processor architecture on the overall execution.

*2) Results & Analysis:* As shown in Figure 2, the tightly-coupled architecture achieves similar performance when compared with HW in most of the benchmarks. In particular, it is found that the execution of MM, FIR and KM on TIGHTLY-

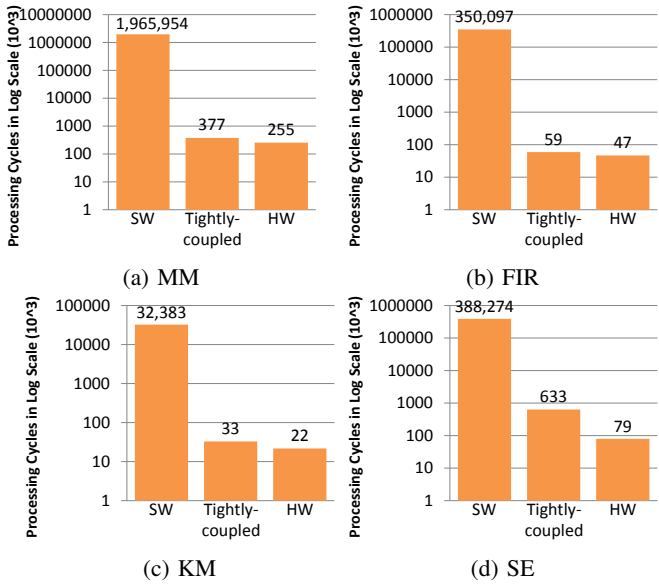

Fig. 2: The performance of SW versus TIGHTLY-COUPLED versus HW.

TABLE V: Resource consumption and maximum frequency of the proposed soft processor versus ORCA RISC-V core.

| Designs | Slice Registers | | Slice LUTs | | Block RAM | | Max. Freq. |
|---|---|---|---|---|---|---|---|
| Proposed Soft Processor | 334 | $\sim 0\%$ | 1279 | $2\%$ | 6 | $4\%$ | 147.929 MHz |
| ORCA RISC-V core | 615 | $\sim 0\%$ | 1438 | $2\%$ | 1 | $\sim 0\%$ | 199.322 MHz |

TABLE VI: Processing cycles and the execution latency of the four benchmarks on the proposed soft processor and ORCA RISC-V core.

| Designs | Benchmarks | # of Cycles | Latency |
|---|---|---|---|
| Proposed Soft Processor | MM | 1965954155 | 13.29 s |
| | FIR | 350096784 | 2.37 s |
| | KM | 32382531 | 0.22 s |
| | SE | 388273610 | 2.62 s |
| ORCA RISC-V core | MM | 2868605367 | 14.39 s |
| | FIR | 503309228 | 2.53 s |
| | KM | 46286121 | 0.23 s |
| | SE | 566543126 | 2.84 s |

COUPLED exhibits comparable performance to that on HW. The main reason is that these applications only need the soft processor to handle the loop control which takes a small amount of time while they rely on the FPGA accelerators to handle the most time-consuming computing.

In the case of SE, however, the number of processing cycles needed in TIGHTLY-COUPLED is significantly more than that in HW. Such performance discrepancy is mainly due to the following two reasons. First, the boundary conditions in SE, i.e. the edge pixels, could not be covered by the auxiliary architecture for acceleration. Therefore a relatively large amount of elements (516 in this case) had to be handled by the soft processor. This would undoubtedly increase the overall execution latency. Second, SE needed the soft processor to perform a large amount of multiplication to calculate the correct memory location for a particular pixel and the entire execution latency will be lengthened consequently, especially when `RV32I` does not include a multiply instruction. This problem can be alleviated by extending the ISA to `RV32IM` and incorporating a multiplier in the soft processor design, which will be supported in the future as one of the customization parameters in the proposed framework.

### B. Evaluation of the Soft Processor

As mentioned in Section I, although the soft processor is mostly responsible for irregular data processing and providing controls over the accelerator, it is still important to have the processor maintaining sufficient efficiency while keeping the area consumption minimal.

*1) Experimental Setup:* In order to study the efficiency of the proposed soft processor, we compared the proposed 4-stage pipeline with ORCA RISC-V core from VectorBlox Computing Inc. [6]. ORCA core is known as an optimized RISC-V design on FPGA that implements a 4-stage pipeline, which is sharing similar design methodology with our proposed soft processor.

In this experiment, we first obtained the maximum frequencies supported by both processors as well as their resource consumptions by synthesising the designs in Xilinx ISE 14.3. Then the benchmarks in Table III were executed using RTL simulator to obtain the processing cycles and subsequently to calculate the execution latency. Note the device chosen for this experiment was Artix 7 xc7a100t-1csg324.

*2) Results & Analysis:* Table V and Table VI display the resource consumption and the performance of the proposed soft processor versus ORCA core. The percentage values in Table V are relative to available resources of the targeted FPGA device.

From these tables, we can see that the proposed soft processor typically occupies less area while at the same time providing slightly higher processing speed. The only resource that the proposed soft processor consumes more than that of ORCA core is the on-chip block RAM. This is mainly due to the existence of the IMEM and DMEM which are inferred as block RAM in the proposed processor. Such memory modules do not exist in ORCA core and hence that explains the discrepancy in block RAM consumption.

### C. Portability and Compatibility Considerations

One of the major advantages of FPGA overlay is to raise the device abstraction level by providing a virtual layer that is conceptually located between user applications and physical

FPGA. Therefore, to integrate the soft processor within the same overlay framework of the accelerator, the processor must be able to support cross-vendors and cross-platforms FPGAs to ensure the device portability and compatibility.

In view of this, we have designed the processor in a generic manner such that the 4-stage pipeline design can be synthesized across various platforms ranging from Spartan-3 to Virtex-7 and Cyclone IV to V. Table VII displays the resource consumption and the maximum frequency of the soft processor implementations on both the high-end and low-end FPGA devices. Again, the percentage values in the table is relative to available resources on the target FPGA device.

## IV. Related Work

In academia, various forms of FPGA overlay have been developed to provide an efficient trade-off between flexibility of software and hardware acceleration for computationally intensive applications. Research works such as ZUMA [7], QUKU [8], or QuickDough [1], [2] have demonstrated the benefits of overlay by improving designers' productivity while maintaining excellent performance.

In spite of this, research that focuses on the integration between the processor and accelerator remains uncommon. The coupling between these two has not been strictly defined or specified. Therefore in many of the existing overlay works, diverse choice of soft/ hard processors [9], [10], [2] are used and the integration between the processor and accelerator varies from one work to another.

The closest work that is designed to resolve the above coupling problem is ADRES [11]. Mei et al. proposed an architecture that contains a VLIW processor tightly-coupled with a coarse-grained reconfigurable matrix. By integrating these two entities together, substantial resource sharing and reduced communication can be obtained. However, it is noticed that ADRES is focusing on enhancing the performance of the entire architecture. The design portability and compatibility of the ADRES framework and the resource consumption of the processor on FPGA are not the major concerns in their work.

The work on soft processors, on the other hand, is extensive both in the industry and academia. Commercial cores such as MicroBlaze [12] and Nios II [13] are the most commonly used soft processors on FPGA. However, design portability of these cores is relatively limited since they are generally restricted to their own platform, and as a result are less favourable for deployment in FPGA overlay.

In academia, many research works on soft processors concentrate on the influence of underlying FPGA architecture. In particular, the processor architecture that is best suitable for the underlying structures of FPGA is comprehensively studied so as to maximize the performance of soft processors [14], [15].

In addition, soft vector processors [16], [17], [18], soft VLIW processors [19], [20], multi-thread soft processors [21], [22] and application-specific soft processors [23], [24] are also extensively studied and developed on FPGA to analyze the performance and to demonstrate the benefits of soft processors.

Although the above cores are optimized for performance by customizing the processors architectures according to the underlying FPGA structures, design portability and resource consumption are not the primary concerns in these works and therefore porting these soft cores onto other platforms as part of the FPGA overlay framework could require tremendous amount of efforts.

Some existing open source soft processors do focus on area optimization and portability such as Plasma [25] and MB-LITE [26]. They are light-weight implementations that can be ported to different platforms in a relatively efficient manner. However, as they are based on MIPS [27] and MicroBlaze ISA respectively, they suffer from patent and trademark issues when being employed commercially.

Moreover, there exists some lightweight `RV32I` designs such as zscale [28], GRVI [29] or ORCA [6] which are similar to the proposed processor. Zscale is a single-issue 3-stage pipeline that is suited for embedded systems. However, its design is not optimized for soft processor implementation and therefore zscale is less desirable for FPGA overlay framework.

GRVI [29] core, on the other hand, is an efficient, FPGA-optimized 3-stage pipeline implementation that is specifically designed for Phalanx framework. It consumes 320 LUTS and runs at $375\,\text{MHz}$ on Kintex Ultrascale-2 FPGA. Compared with GRVI, the proposed processor puts emphasis on design portability and compatibility as well as architectural extension to the tightly-coupled accelerator.

## V. Conclusion

This paper has presented an open-source soft processor that is tightly-coupled with an FPGA accelerator to become part of an FPGA overlay framework. RISC-V is chosen as the instruction set for its openness and portability, and the soft processor is designed as a 4-stage pipeline to minimize the resource consumption. Experiments show that, when compared with existing design, the proposed soft processor is small and efficient, and the tightly-coupled architecture can provide a unified programming model for software designers while at the same time maintaining certain performance. Furthermore, the processor is generically implemented so as to promote design portability and compatibility across different FPGA platforms. It is anticipated that the proposed architecture can eventually improve the portability of FPGA overlays, and promote the use of FPGA among software developers.

The soft processor is available at https://github.com/hku-casr/riscv-overlay.

## Acknowledgment

This work was supported in part by the Research Grants Council of Hong Kong project ECS 720012E and the Croucher Innovation Award 2013.

TABLE VII: Resource consumption and maximum frequency of the proposed soft processor on high-end and low-end FPGA devices across various generations.

| Devices | Slice Flip Flops | | 4 Input LUTs | | RAMB16s | | Max. Freq. |
|---|---|---|---|---|---|---|---|
| Spartan3 xc3s50-5vq100 | 304 | 19 % | 1461 | 95 % | 2 | 50 % | 76.127 MHz |
| Virtex4 xc4vfx140-11ff1517 | 248 | ∼ 0 % | 1344 | 1 % | 4 | ∼ 0 % | 157.351 MHz |
| **Devices** | **Slice Registers** | | **Slice LUTs** | | **Block RAM** | | **Max. Freq.** |
| Virtex5 xc5vlx30-1ff324 | 280 | 1 % | 1081 | 5 % | 2 | 6 % | 184.735 MHz |
| Virtex5 xc5vlx155t-3ff1738 | 266 | ∼ 0 % | 1070 | 1 % | 2 | ∼ 0 % | 231.913 MHz |
| Spartan6 xc6slx4-2tqg144 | 369 | 7 % | 1296 | 61 % | 10 | 83 % | 88.355 MHz |
| Virtex6 xc6vhx380t-3ff1923 | 321 | ∼ 0 % | 1263 | ∼ 0 % | 6 | ∼ 0 % | 203.198 MHz |
| Artix7 xc7a100t-1csg324 | 334 | ∼ 0 % | 1279 | 2 % | 6 | 4 % | 147.929 MHz |
| Virtex7 xc7vx690t-3ffg1927 | 318 | ∼ 0 % | 1280 | ∼ 0 % | 6 | ∼ 0 % | 268.666 MHz |
| **Devices** | **Logic Registers** | | **Logic Elements** | | **Memory Bits** | | **Max. Freq.** |
| Cyclone IV EP4CE6F17C8 | 284 | 4 % | 1890 | 30 % | 165 888 kbit | 60 % | 68.43 MHz |
| Cyclone IV EP4CGX75DF27C6 | 284 | 3 % | 1890 | 3 % | 165 888 kbit | 4 % | 93.07 MHz |
| **Devices** | **Logic Registers** | | **ALM** | | **Memory Bits** | | **Max. Freq.** |
| Cyclone V 5CEBA2F17C8 | 284 | 2 % | 784 | 8 % | 165 888 kbit | 9 % | 73.5 MHz |
| Cyclone V 5CGXFC7D6F31C6 | 284 | 1 % | 791 | 1 % | 165 888 kbit | 2 % | 102.76 MHz |